\documentclass{sig-alternate}
\usepackage{mathptmx} 

\usepackage{fancyhdr}
\usepackage[normalem]{ulem}
\usepackage[hyphens]{url}
\usepackage[sort,nocompress]{cite}
\usepackage[final]{microtype}
\usepackage{flushend}
\usepackage[numbers,sort&compress,square]{natbib}
\usepackage{amsmath,amssymb,amsfonts}
\usepackage{algorithmic}
\usepackage{graphicx}
\usepackage{textcomp}
\usepackage{xcolor}
\usepackage{fancyhdr}
\usepackage[hyphens]{url}
\usepackage{xspace}
\usepackage{booktabs}
\usepackage{setspace}
\usepackage{enumitem}
\usepackage{etoolbox}
\usepackage{makecell}
\usepackage[bookmarks=true,breaklinks=true,letterpaper=true,colorlinks,citecolor=blue,linkcolor=blue,urlcolor=blue]{hyperref}

\usepackage{tikz}
\usepackage{soul}
\usepackage{multirow}
\usepackage{graphicx}
\usepackage{array}
\usepackage{footmisc}
\usepackage{setspace}

\newcolumntype{C}[1]{>{\centering\let\newline\\\arraybackslash\hspace{0pt}}m{#1}}
            
\def\BibTeX{{\rm B\kern-.05em{\sc i\kern-.025em b}\kern-.08em
    T\kern-.1667em\lower.7ex\hbox{E}\kern-.125emX}}

\newcommand*\circledb[1]{\tikz[baseline=(char.base)]{
            \node[shape=circle,fill,inner sep=0.5pt] (char) {\textcolor{white}{#1}};}}

\fancypagestyle{firstpage}{
  \fancyhf{}
  
  \fancyfoot[C]{\thepage}
}

\author{\vspace{-25pt}\\
{Georgia Antoniou$^{\ddagger}$\textsuperscript{$\star$}} \quad \quad \quad \quad \quad
{Haris Volos$^{\ddagger}$} \quad \quad \quad \quad \quad
{Davide B. Bartolini$^{\S}$} 
\vspace{7pt}\\
{Tom Rollet$^{\S}$} \quad \quad \quad \quad \quad~~~ 
{Yiannakis Sazeides$^{\ddagger}$} \quad \quad \quad \quad 
{Jawad Haj Yahya$^{\S}$} 
\vspace{10pt}\\
{\fontsize{10}{11}\selectfont
$^{\S}$\textit{Huawei Technologies - Zurich Research Center}\quad \quad \quad
$^{\ddagger}$\textit{University of Cyprus} 
}
\vspace{0pt}}

\newcommand{\techL}{AgilePkgC\xspace}
\newcommand{\tech}{APC\xspace}
\newcommand{\CAgile}{$PC1A$\xspace}

\newcommand\hj[1]{{\color{black}{#1}}}
\newcommand\dbb[1]{{\color{black}{#1}}}
\newcommand\hv[1]{{\color{black}{#1}}}
\newcommand\ga[1]{{\color{black}{#1}}}
\newcommand\ys[1]{{\color{black}{#1}}}

\newcommand{\hp}[2][pink]{{%
    #2}%
}

\newcommand{\rG}[1]{%
{\color{black}#1}%
}

\newcommand\asm[0]{{IO Standby Mode}\xspace}
\newcommand\ASM[0]{IOSM\xspace}
\newcommand\ccsm[0]{{CHA, LLC, and Mesh Retention}\xspace}
\newcommand\CCSM[0]{CLMR\xspace}
\newcommand\CLMR[0]{CLMR\xspace}

\title{\vspace{-13pt}\techL: An Agile System Idle State Architecture \\for Energy Proportional Datacenter Servers\vspace{-28pt}}

\begin{document}

\maketitle
\thispagestyle{firstpage}
\pagestyle{plain}

\AtEndEnvironment{thebibliography}{
\bibitem{B1}  Barroso, Luiz André, Jeffrey Dean, and Urs Holzle. ``Web search for a planet: The Google cluster architecture.'' IEEE micro 2003. 
\bibitem{B2} Jeon, M., He, Y., Elnikety, S., Cox, A.L. and Rixner, S., 2013, April. ``Adaptive parallelism for web search''. EuroSys 2013.
\bibitem{B3} D. Wong and M. Annavaram. ``Knightshift: Scaling the Energy Proportionality Wall Through Server-level
Heterogeneity''. In MICRO, 2012.
\bibitem{B4} S. Luo, H. Xu, C. Lu, K. Ye, G. Xu, L. Zhang, Y. Ding, J. He, and C. Xu. ``Characterizing Microservice Dependency and Performance: Alibaba Trace Analysis''. In SoCC, 2021.
\bibitem{spark} Github, ``Spark-Bench''
\bibitem{hive} Apache , ``Apache Hive''

}
\begingroup\renewcommand\thefootnote{$\star$}
\footnotetext{This work was done while Georgia Antoniou was intern at Huawei Technologies - Zurich Research Center.}
\endgroup

\begin{abstract}
Modern user-facing applications deployed in datacenters use a distributed system architecture that exacerbates the latency requirements of their constituent microservices ($30$--$250{\mu}s$).
Existing CPU power-saving techniques degrade the performance of these applications due to the long transition latency (order of $100{\mu}s$) to wake up from a deep CPU idle state (C-state).
For this reason, server vendors recommend only enabling shallow \emph{core C-states} (e.g., $CC1$) for idle CPU cores, thus preventing the system from entering deep \emph{package C-states} (e.g., $PC6$) when all CPU cores are idle.
This choice, however, impairs server energy proportionality since power-hungry resources (e.g., IOs, uncore, DRAM) remain active even when there is no active core to use them.
As we show, it is common for all cores to be idle due to the low average utilization (e.g., $5-20\%$) of datacenter servers running user-facing applications. 

We propose to reap this opportunity with \techL (\tech), a new package C-state architecture that improves the energy proportionality of server processors running latency-critical applications. 
\tech implements $PC1A$ (package $C1$ agile), a new deep package C-state that a system can enter once all cores are in a shallow C-state (i.e., $CC1$) and has a nanosecond-scale transition latency. $PC1A$ is based on four key techniques.
First, a hardware-based \emph{agile power management unit} (APMU) rapidly detects when all cores enter a shallow core C-state ($CC1$) and trigger the system-level power savings control flow. 
Second, an \emph{\asm}~(\ASM) that places IO interfaces (e.g., PCIe, DMI, UPI, DRAM) in shallow (nanosecond-scale transition latency) low-power modes.
Third, a \emph{CLM Retention}~(\CLMR) rapidly reduces the CLM (Cache-and-home-agent, Last-level-cache, and Mesh network-on-chip) domain's voltage to its retention level, drastically reducing its power consumption.
Fourth, \tech keeps all system PLLs active in $PC1A$ to allow nanosecond-scale exit latency by avoiding PLLs' re-locking overhead. 

Combining these techniques enables significant power savings while requiring less than $200$ns transition latency, ${>}250\times$ faster than existing deep package C-states (e.g., $PC6$), making $PC1A$ practical for datacenter servers.
Our evaluation using Intel Skylake-based server  
shows that \tech reduces the energy consumption of Memcached by up to $41\%$ ($25\%$ on average) with ${<}0.1\%$ performance degradation. 
\tech provides similar benefits for other representative workloads.
\end{abstract}

\section{Introduction}
\label{sec:intro}
The development of cloud applications running in datacenters is increasingly moving away from a monolithic to microservice software architecture to facilitate productivity~\cite{Fowler2014,dmitry2014micro}. This comes at the expense of application performance becoming more vulnerable to events that result in ``killer'' microsecond scale idleness~\cite{barroso2017attack}. This is acute for user-facing applications with tight tail-latency requirements whereby serving a user query typically consists of executing numerous interacting microservices that explicitly communicate with each other~\cite{barroso2017attack,barroso2018datacenter,prekas2017zygos}. The communication latency limits the time available to execute a microservice and magnifies the impact of microsecond scale idleness (e.g., events related to NVM, main memory access, and power management)~\cite{barroso2017attack,chou2019mudpm,cho2018taming}. 
This is further compounded by the dynamics of user-facing applications' unpredictable and bursty load ~\cite{chou2016dynsleep,chou2019mudpm,meisner2009powernap}.
As a result, each microservice needs to operate under a tight (i.e., tens to hundreds of $\mu$s) latency requirement~\cite{zhan2016carb,chou2019mudpm}. 

One widely used method to ensure that microservices, and hence overall applications, meet their performance target is to execute them on servers that have low average utilization ($5$--$20\%$)~\cite{lo2014towards,B1,B2,B3,B4,iorgulescu2018perfiso}, leading to a busy/idle execution pattern~\cite{chou2016dynsleep,chou2019mudpm,meisner2009powernap} where cores are frequently idle.
Ideally, each core should enter a low-power \emph{core C-state} whenever it is idle, and the entire system 
should transition to a low-power \emph{package C-state} whenever \emph{all cores} are idle. However, the situation in modern datacenters is quite different.

\begin{table}[h]
\vspace{-15pt}
\centering
\caption{Power across existing package C-states and our new $PC1A$ for our baseline server (details in \autoref{sec:methodology}).}
\label{tab:c-states}
\resizebox{\columnwidth}{!}{%
\begin{tabular}{lllrlr}
{\bf Package} &{\bf / cores C-state\footref{note:pc_cc}} &
\textbf{Latency}\footref{note:latency} &
{\bf SoC} & \multicolumn{2}{l}{{\bf + DRAM power}} \\
\cmidrule(lr){1-2} \cmidrule(lr){3-3} \cmidrule(lr){4-6}
$PC0$  &/ ${\geq}1 $~$CC0$  & $0$ns          & ${\leq}85$W &+ $7$W & ${\leq}~92.0$W \\
$PC0_{\dbb{idle}}$  &/ 10 $CC1$  & $0$ns          & $44$W &+ $5.5$W  &=~$49.5$W \\
$PC6$  &/ 10 $CC6$  & ${>}50\mu$s    & $12$W &+ $0.5$W  &=~$12.5$W \\
$\mathbf{PC1A}$ &/ {\bf 10} $\mathbf{CC1}$ & $\mathbf{<200}${\bf ns}      & $27.5$W &+ $1.6$W  &=~$\mathbf{29.1}${\bf W}
\end{tabular}%
}
\vspace{-5pt}
\end{table}

\autoref{tab:c-states} reports power consumption and transition latency%
\footnote{We report worst-case entry+exit latency for the package C-state to open the path to memory; the overall latency to resume execution may be higher and include core C-state latency~\cite{gough2015cpu,haj2018power}.\label{note:latency}}
for the processor system-on-chip (SoC) and DRAM in a typical server for existing package C-states and our  proposed package C-state, $PC1A$ (introduced in \autoref{sec:technique}).
If any core is active (i.e., $CC0$ C-state%
\footnote{We refer to core C-states as $CCx$ and package C-states as $PCx$; higher values of $x$ indicate deeper, lower-power C-states.\label{note:pc_cc}}%
), the system is also active (i.e., $PC0$ package C-state).
A core can enter a deeper C-state (e.g., $CC1$, $CC6$) when it is idle, and similarly, the system can enter a deeper package C-state (e.g., $PC6$) when all cores reside at the same time in a deep core C-state ($CC6$).
However, the high transition latency imposed by $CC6$ (and, subsequently, $PC6$), coupled with short and unpredictable request arrivals, severely reduces the usefulness of these deep C-states in datacenter servers.
Server vendors \emph{recommend disabling} deep core C-states in datacenters to prevent response-time degradation~\cite{cisco_cstates,dell_cstates,lenovo_cstates,intel_idle}.
Consequently, existing package C-states can never be entered even when all cores are idle in $CC1$
(e.g., Intel modern servers can only enter $PC6$ if all cores are in $CC6$)~\cite{gough2015cpu,intel_2nd_gen_xeon_datatsheet}.
This scenario in datacenter servers results in significant power waste as the uncore and other shared components (e.g., DRAM) fail to enter any low-power state  when all cores are idle.

A seminal work by Google that discusses latency-critical applications states \mbox{\cite{lo2014towards}}: ``Modern servers are not energy proportional: they operate at \emph{peak} energy efficiency when they are fully utilized but have much lower efficiencies at lower utilizations''. The utilization of servers running latency-critical applications is typically $5\%$--$20\%$ to meet target tail latency requirements, as reported by multiple works from industry and academia \mbox{\cite{lo2014towards,B1,B2,B3,B4}}. For example, recently, Alibaba reported that the utilization of servers running latency-critical applications is typically $10\%$ \cite{B4}. Therefore, to improve the energy proportionality of servers running latency-critical microservice-based applications, it is crucial to address the more inefficient servers' operating points, namely the \emph{low utilization, which is the focus of our study}.

Prior work (reviewed in \autoref{sec:related}) proposes various management techniques to mitigate the inability of datacenter processors to leverage deep C-states effectively.
In contrast, our \textbf{goal} is to directly address the root cause of the inefficiency, namely the high transition latency (tens of $\mu$s; see \autoref{tab:c-states}) of deep package C-states.
\textbf{To this end}, we propose \emph{\techL(\tech)}: 
a new package C-state architecture to improve the energy proportionality of server processors running latency-critical applications.
\tech introduces $PC1A$: a low-power package C-state with nanosecond-scale transition latency that the system can enter as soon as all cores enter shallow C-states (e.g., $CC1$, rather than after all cores enter deeper C-states, e.g., $CC6$, which are unreachable as they are normally disabled in server systems).
A low-latency package C-state is crucial since periods of whole-system idleness (i.e., all cores idle) are even shorter and more unpredictable than idle periods of individual cores.

\tech leverages \emph{four} key power management techniques that differentiate $PC1A$ from existing package C-states.
%
1) A hardware-based \emph{agile power management unit} (APMU) to rapidly detect when all cores enter a shallow core C-state ($CC1$) and trigger a system-level power savings flow. 
2) An \emph{\asm}~(\ASM)  that places IO interfaces (e.g., PCIe, DMI, UPI, DRAM) in shallow (nanosecond-scale transition latency) low-power modes.
3) A \emph{CLM Retention}~(\CLMR) that leverages the fast integrated voltage regulator~\cite{burton2014fivr,nalamalpu2015broadwell} to rapidly reduce the CLM (Cache-and-home-agent, Last-level-cache, and Mesh network-on-chip) domain's voltage to its retention level, drastically reducing CLM's power consumption.
4) \tech keeps all system PLLs active in $PC1A$ to allow nanosecond-scale exit latency by avoiding PLLs' re-locking latency (a few microseconds). This approach significantly reduces transition latency at a minimal power cost, thanks to modern all-digital PLLs' energy efficiency~\cite{fayneh20164}. 

%
Our evaluation using Intel Skylake-based server  
shows that \tech reduces the energy consumption of Memcached \cite{memcached} by up to $41\%$ ($25\%$ on average) with ${<}0.1\%$ performance degradation. 
\tech provides similar benefits for other representative workloads.
\tech's new package C-states, $PC1A$, exhibits more than $250\times$ shorter transition latency than the existing deep package C-state $PC6$.

While we demonstrate \tech potential for Intel servers, which account for more than $80\%$ of the entire server processor market~\cite{intel_amd_marketshare}, our proposed techniques are general, hence applicable to other server processor architectures.

In summary, this work makes the following \textbf{contributions}:
\begin{itemize}[leftmargin=*,nolistsep]
\item \tech is the first practical package C-state design targeting the killer microseconds problem in datacenter servers running latency-critical applications. 
%
\item \tech introduces the $PC1A$ low-power package C-state that a system can enter once all cores enter a shallow C-state (i.e., CC1). \item \tech improves existing deep package C-states by drastically reducing their transition latency (${>}250\times$) while retaining a significant fraction of their power savings.

%
%
\item Our evaluation 
shows  that  \tech  reduces  the energy consumption of Memcached by up to $41\%$  with  less than $0.1\%$ performance degradation. \tech achieves similar gains for other representative workloads.
\end{itemize}

\vspace{-5pt}
\section{Motivation}
\label{sec:motivation}

\newcommand\CZeroW[0]{$4$W\xspace}
\newcommand\COneW[0]{$1.44$W\xspace}
\newcommand\COneT[0]{$2\mu$s\xspace}
\newcommand\CSixW[0]{$0.1$W\xspace}

Modern servers running latency-critical applications are stuck in $PC0$ (i.e., active package C-state) and never enter $PC6$, because $CC6$ is disabled in these systems~\cite{cisco_cstates,dell_cstates,lenovo_cstates}.
A major consequence of this is that the server experiences high power consumption from the uncore components in the processor SoC (e.g., last-level-cache, IO interfaces) and DRAM, which are always active~\cite{gough2015cpu}.
Our measurements (see \autoref{sec:methodology}) of an idle system (all cores in $CC1$) show that uncore \& DRAM power consumption accounts for more than $65\%$ of the SoC \& DRAM power consumption.

Adding a deep agile package C-state $PC1A$ that 1) has a sub-microsecond transition time and 2) only requires cores to enter $CC1$ would significantly improve energy proportionality for servers by drastically reducing uncore and DRAM power consumption when all cores are idle.
\autoref{eq:savings} estimates the power savings that $PC1A$ C-state could bring.

\begin{figure}[h!]
 \vspace{-1.5em}
\begin{align}
P_{baseline} &= R_{PC0} \times P_{PC0} + R_{PC0_{idle}} \times P_{PC0_{idle}} \nonumber\\
\%P_{savings} &= R_{PC1A} \times \left( P_{PC0_{idle}} - P_{PC1A} \right)  /  P_{baseline} \label{eq:savings}
\end{align}
\vskip-1.em
\end{figure}
$P_{baseline}$ is the overall, SoC \& DRAM, power of a current server measured as the sum of the power while the system has at least one core in $CC0$  and when all cores are idle in $CC1$ (i.e., $P_{PC0}$ and $P_{PC0_{idle}}$) weighted by their respective state residencies $R_{PC0}$ and $R_{PC0_{idle}}$. We can obtain the savings of $PC1A$ from \autoref{eq:savings} by using the power of the new proposed state $P_{PC1A}$ (shown in \autoref{tab:c-states} and derived in Sec.~\ref{sec:impl}) and assuming that the fraction of time a server will spend in $PC1A$ is the same as the time the baseline spends in $PC0_{idle}$ (i.e., $R_{PC1A}=R_{PC0_{idle}}$).



For example, we consider running a key-value store  workload (e.g., Memcached~\cite{jose2011memcached})
\begin{figure*}[!hpt]
	\vspace{-18pt}
    \centering
	\includegraphics[trim=0.5cm 0.5cm 0.5cm 0.6cm, clip=true,width=0.96\linewidth,keepaspectratio]{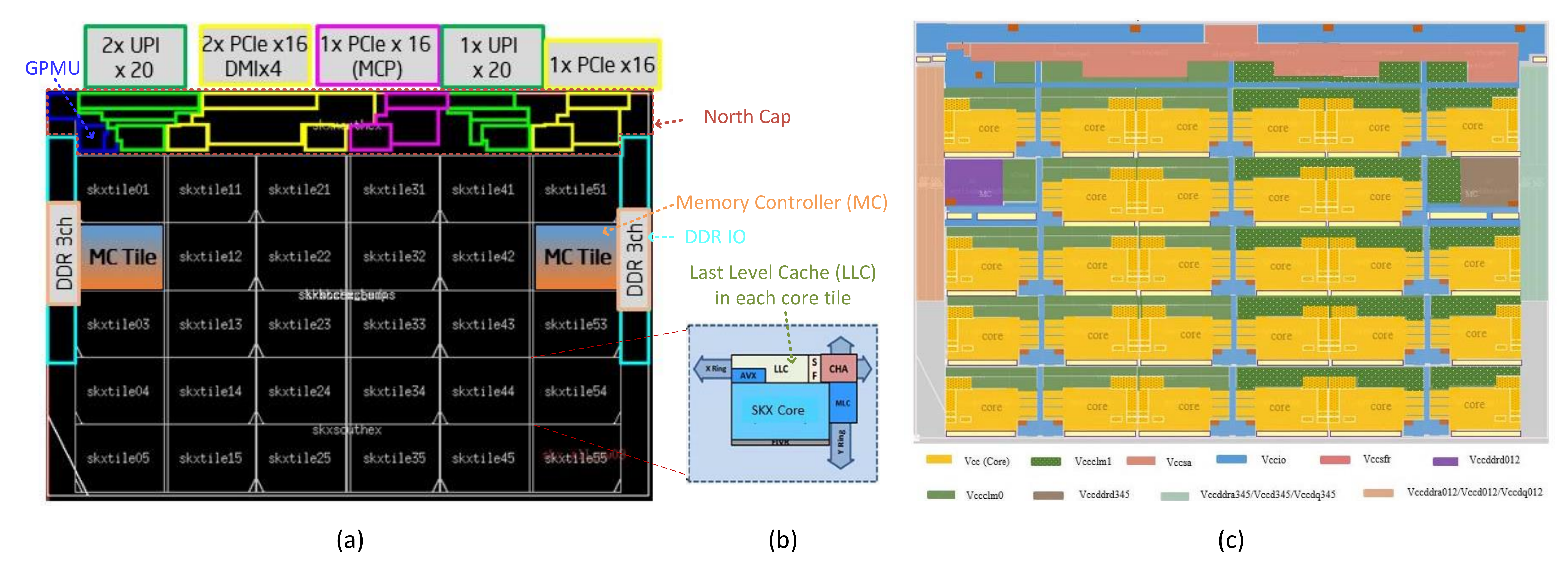}
	\vspace{-11pt}
	\caption{Skylake server (SKX) architecture (a) SKX tiled floorplan (b) SKX tile (c) SKX voltage domains~\cite{tam2018skylake}.}
	\label{fig:skx_floorplan}
 	\vspace{-14pt}
\end{figure*}
on a processor with $10$ cores. Our experimental analysis (see \autoref{sec:methodology}) reveals that all cores are simultaneously in $CC1$ state for ${\sim}57\%$ and ${\sim}39\%$ of the time at $5\%$ and $10$\% load, respectively. Plugging power numbers from our evaluation (see \autoref{sec:methodology} and \autoref{sec:evaluation}) in the power model shows that  placing the system into $PC1A$ when all cores are at $CC1$ can save $23\%$ and $17\%$ for a $5\%$ and $10\%$ loaded system, respectively. For an idle server, i.e., during times with no tasks assigned to the server, $R_{PC0}=0\%$ and $R_{PC0_{idle}}=100\%$, and \autoref{eq:savings} is simplified to $1 - P_{PC1A} / P_{PC0_{idle}}$; hence $PC1A$ can reduce power consumption by ${\sim}41\%$.


\section{Background}
\label{sec:bg}


\autoref{fig:skx_floorplan}(a) shows the floorplan for an Intel Skylake Xeon server processor (SKX), consisting of three major building blocks: the mesh tiles, north-cap, and DDR IOs (PHYs).
SKX uses a mesh network-on-chip to connect cores, memory controllers (MC), and IO controllers (North Cap)~\cite{tam2018skylake,Skylake_die_server,mesh_interconnect}.

\noindent\textbf{Core tiles.}
The largest area contributor to the entire SoC area are the core tiles (\autoref{fig:skx_floorplan}(b)). Each of which contains 1) all core domain (CPU core, AVX extension, and private caches) and 2) a portion of the uncore domain (caching-and-home-agent (CHA), last-level-cache (LLC), and a snoop filter (SF))~\cite{tam2018skylake}. 


\noindent\textbf{North-Cap.}
The top portion of the SoC die is called the {\it north-cap}~\cite{tam2018skylake,Skylake_die_server}. It consists of the high-speed IO (PCIe, UPI, and DMI) controllers and PHYs, serial ports, fuse unit, clock reference generator unit, and the firmware-based global power management Unit (GPMU).

\noindent\textbf{Power Delivery Network (PDN).}
The PDN is the SoC subsystem responsible for providing stable voltage to all the processor domains \cite{burton2014fivr,nalamalpu2015broadwell,tam2018skylake,icelake2020,haj2020flexwatts}.
\autoref{fig:skx_floorplan}(c) shows the organization of the SoC into voltage domains.
SKX implements~\cite{tam2018skylake} nine primary voltage domains generated using  a FIVR (fully integrated voltage regulator ~\cite{burton2014fivr,nalamalpu2015broadwell,tam2018skylake,icelake2020}) or MBVR (motherboard voltage regulator~\cite{rotem2011power,fayneh20164,haj2019comprehensive}). For example, each core has a dedicated FIVR (Vcc core), and the CLM (CHA, LLC, mesh interconnect) has two FIVRs (Vccclm0 and Vccclm1); IO controllers and PHYs use MBVR (Vccsa and Vccio, respectively) \cite{tam2018skylake}.

\noindent\textbf{Clock Distribution Network (CDN).}
A CDN distributes the signals from a common point (e.g., clock generator) to all the elements in the system that need it. 
Modern processors use an all-digital phase-locked loop (ADPLL) to generate the CPU core clock~\cite{tam2018skylake}. An ADPLL maintains high performance with significantly less power as compared to conventional PLLs \cite{fayneh20164}. 
SKX system uses multiple PLLs: a PLL per core \cite{tam2018skylake}, a PLL per each high-speed IO (i.e., PCIe, DMI, and UPI controller) \cite{intel_2nd_gen_xeon_datatsheet}, one PLL for the CLM domain \cite{tam2018skylake}, and one PLL for the global power management unit \cite{Skylake_die_server}.

\subsection{Power Management States}
\label{sec:pm_states}

Power management states reduce power consumption while the system or part of it is idle. Modern processors support multiple power states such as Core C-states, IO link-state (L-state), DRAM power mode, and Package C-state. 

\noindent\textbf{Core C-states (CCx).}
Power saving states enable cores to reduce their power consumption during idle periods. We refer to core C-states as $CCx$; $CC0$ is the active state, and higher values of $x$ correspond to deeper C-states, lower power, and higher transition latency.
For example, the Intel Skylake architecture offers four core C-states: $CC0$, $CC1$, $CC1E$, and $CC6$~\cite{gough2015cpu,haj2018power,schone2015wake}. 
While C-states reduce power, a core cannot be utilized to execute instructions during the entry/exit to/from a C-state. For example, it is estimated that $CC6$ requires $133${\textmu}s transition time  \cite{intel_idle,CPU_idle}. As a result, entry-exit latencies can degrade the performance of services that have microseconds processing latency, such as in user-facing applications~\cite{jose2011memcached}.

\noindent{\bf IO L-states (Lx).}
High-speed IOs (Links) support power states that provide similar performance/power trade-offs to core C-states~\cite{gough2015cpu}.
While specific power states differ based on the type of link, the high-level concepts we describe here are similar.
$L0$ is the active state, providing maximum bandwidth and minimum latency.
$L0s$ is a standby state, during which a subset of the IO lanes are asleep and not actively transmitting data. The reference clock and internal PLLs are kept active to allow fast wakeup (typically ${<}64ns$~\cite{l0s_entry_lat,icelake2020L0s,gough2015cpu}) while providing significant (up to ${\sim}50\%$ of $L0$) power savings.
$L0p$ is  similar to $L0s$ state, but a subset of the data lanes remain awake (typically half). Bandwidth is reduced, and latency for transmitting data increases. $L0p$ provides up to ${\sim}25\%$ lower power than $L0$ with faster exit latency than $L0s$ (typically ${\sim}10ns$). The IO link-layer autonomously handles the entry to $L0s$/$L0p$ states (no OS/driver interactions) once the IO link is idle~\cite{gough2015cpu}.
$L1$ is a power-off state, meaning that the link must be retrained, and PLLs must be switched on to resume link communication. $L1$ provides higher power saving than $L0s$ and $L0p$ but requires a longer transition latency (several microseconds).

\noindent{\bf DRAM Power Saving Techniques.}
Modern systems implement two main DRAM power-saving techniques: \emph{CKE modes} and \emph{self-refresh} \cite{gough2015cpu,appuswamy2015scaling,david2011memory,malladi2012rethinking}.

\emph{CKE modes:} CKE (clock enable) is a clock signal the memory-controller (MC) sends to the DRAM device. When the MC turns-off the CKE signal, the DRAM can enter low power modes.
There are two main types of CKE power-modes in DDR4: 1) Active Power Down (APD), which keeps memory pages open and the row buffer powered on, and 2) Pre-charged Power Down (PPD), which closes memory pages and powers down the row buffer.
The granularity of CKE modes is per rank and it is considered a relatively quick technique (independent of the power mode used), with nanosecond-scale transition latency ($10$ns -- $30$ns) and significant power savings (${\geq}50\%$ lower power than active state) \cite{appuswamy2015scaling,david2011memory,malladi2012rethinking}.

\emph{Self-refresh:} In system active state, the MC is responsible to issue the refresh commands to DRAM. To reduce power consumption in MC and DRAM device, DRAM support a self-refresh mode, in which the DRAM is responsible for the refresh process. Once the MC places the DRAM in Self-refresh mode, the power management unit can turn-off the majority of the interface between the SoC and the DRAM \cite{haj2020techniques}. Due to this deep power-down, the exit latency of self-refresh is several microseconds. To minimize the performance impact of self-refresh exit latency, the power management unit of modern processors allow transitions to the self-refresh state only while in a deep idle power state (e.g., package C-states) \cite{gough2015cpu,appuswamy2015scaling,haj2018power}.

\begin{table}[h]
\vspace{-15pt}
\caption{SKX package C-state characteristics and our new $PC1A$ (details in \autoref{sec:technique}).}
\label{tbl:uarch_state_pcx}
\centering
\resizebox{\columnwidth}{!}{%
\begin{tabular}{lllllll}
 \textbf{PCx} & \textbf{Cores in CCx} & \textbf{L3 Cache} & PLLs & \textbf{PCIe/DMI} & \textbf{UPI} & \textbf{DRAM}  \\ 
\cmidrule(lr){1-1} \cmidrule(lr){2-2} \cmidrule(lr){3-3} \cmidrule(lr){4-4} \cmidrule(lr){5-5} \cmidrule(lr){6-6} \cmidrule(lr){7-7}
 PC0 & ${\geq}1$ in CC0 & Accessible & On & L0 & L0 & Available  \\ 
 PC6 & All in CC6 & Retention  & Off & L1 & L1 & Self Refresh\\
 PC1A & All in CC1 & Retention & On & L0s & L0p& CKE off\\
\end{tabular}
}
\vspace{-5pt}
\end{table}

\noindent{\bf Package C-states (PCx).}
Package C-states are responsible for reducing power consumption of the uncore and other system components (e.g., DRAM) once  all cores are idle.
Skylake servers support \emph{three} main package c-states: $PC0$, $PC2$, $PC6$ \cite{schone2015wake,gough2015cpu}.
$PC0$ is the active state, enabled when at least one core is active (i.e., in $CC0$).
$PC2$ is a non-architectural (intermediate) state, 
which is typically used as a transient state between $PC0$ and deeper package C-states.
$PC6$ is a deep package C-state that saves significant uncore power by placing the IOs in low power modes and reducing the CLM voltage to retention, as shown in \autoref{tbl:uarch_state_pcx}. However, $PC6$ has high (${>}50us$, see \autoref{tab:c-states}) transition latency. 
Due to its high transition latency, server vendors typically recommended disabling $PC6$ to prevent response time degradation~\cite{cisco_cstates,dell_cstates,lenovo_cstates}.
%

\begin{figure}[h]
    \centering
	\includegraphics[trim=0.6cm 0.6cm 0.6cm 0.6cm,clip=true,width=1\linewidth,keepaspectratio]{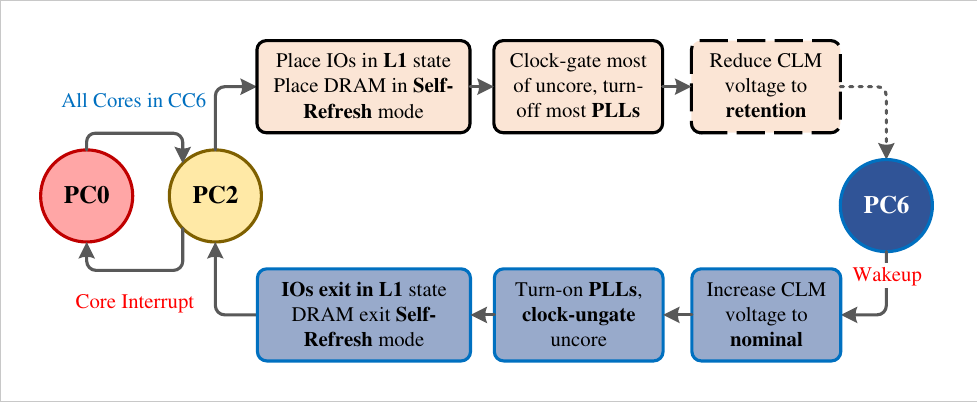}
 	\vspace{-8pt}
	\caption{PC6 entry/exit flow.}
	\label{fig:pc6_c_state_flow}
     \vspace{-5pt}
\end{figure}

\noindent \textbf{PC-state Entry and Exit Flow.}
\autoref{fig:pc6_c_state_flow} illustrates the entry and exit flows for $PC6$~\cite{gough2015cpu, haj2018power, schone2015wake}.
The $PC6$ entry flow starts once all cores reach $CC6$. The flow then moves to the intermediate $PC2$ state. Next, it triggers deep power-saving states in IOs (L1 state) and DRAM (self-refresh mode), clock-gates most of the uncore and turns-off most PLLs.
Then, it reduces the CLM voltage to retention. (see \autoref{tbl:uarch_state_pcx}). When a wake-up event occurs, the system exits from $PC6$ state by reversing the entry flow.
$PC6$ delivers significant power saving, but requires high transition latency (${>}50us$, see \autoref{tab:c-states}).


\vspace{-4pt}
\section{\techL ({\tech}) Architecture}
\label{sec:technique}

\begin{figure}[tp]
\centering
\includegraphics[trim=0.5cm 0.6cm 0.5cm 0.6cm, clip=true,width=\linewidth,keepaspectratio]{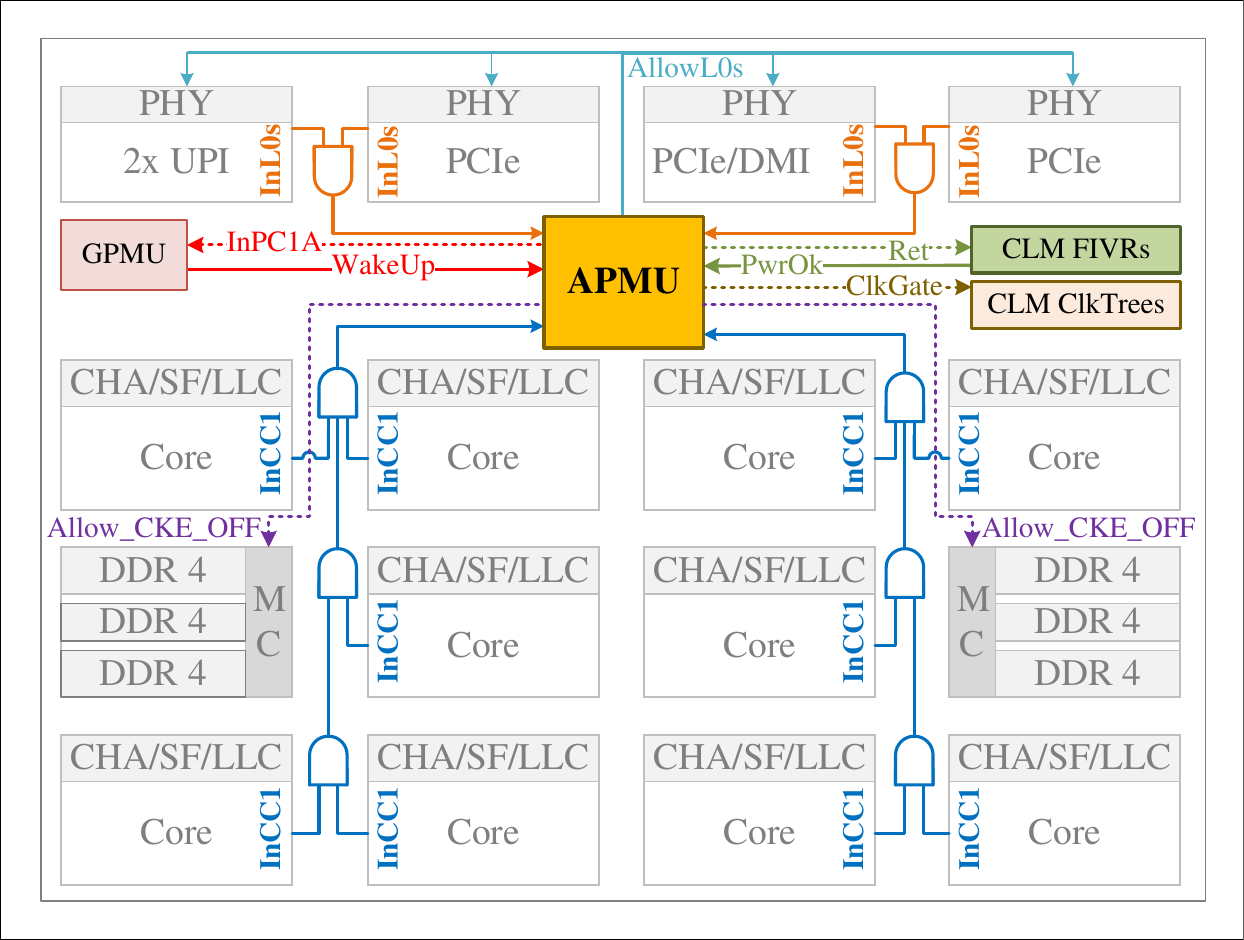}
\\
\vspace{4pt}
\begin{spacing}{0.6}
\scriptsize
APMU / GPMU: agile / global power management unit, %
CLM FIVRs: voltage regulators for the mesh interconnect, %
CHA/SF/LLC: principal uncore components (tiled across cores), %
PHY: device physical layer, %
MC: memory controller
.
\end{spacing}
\caption{Main \tech architecture components (in color).}
\label{fig:pc1_arch}
\vspace{-5pt}
\end{figure}

%
The main \tech components introduced to implement the new $PC1A$ package C-state are shown in \autoref{fig:pc1_arch}.
%
This architecture is based on \emph{three} main components: 
1) the \emph{Agile Power Management Unit} (APMU),
2) the \emph{\asm} (\ASM), and
3) the \emph{\ccsm} (\CCSM), 
discussed in \autoref{sec:powman-flow}, \autoref{sec:ASM}, and \autoref{sec:CCSM},  respectively.
APMU triggers  $PC1A$ system-level power management flow once all cores enter the $CC1$ shallow C-state (see \autoref{tbl:uarch_state_pcx}) and requires additional signals, red in \autoref{fig:pc1_arch}, to interface with the existing, firmware-based global PMU (GPMU).
\ASM enables power saving in the IO domain (i.e., PCIe, DMI, UPI, DRAM) by exploiting IO shallow low-power modes and requires adding specific signals depicted in blue, orange, and purple in \autoref{fig:pc1_arch}.
\CCSM enables power savings in the CLM domain and requires adding two signals to CLM's FIVRs and one to CLM's clock tree, shown in green and brown in \autoref{fig:pc1_arch}.

We first describe the APMU and the $PC1A$ transition flows that it implements, then we describe in detail the \ASM (\autoref{sec:ASM}) and \CCSM (\autoref{sec:CCSM}) components  $PC1A$ uses.  

\vspace{-4pt}
\subsection{Agile Power Management Unit (APMU)}
\label{sec:powman-flow}

\tech introduces APMU to enable system-level power savings by entering $PC1A$ with nanosecond-scale transition latency.
This innovation involves agile coordination of multiple \emph{SoC domains} (e.g., CPU cores, high-speed IOs, CLM, DRAM).
Whereas, rather than trying to enter  domain's deep power states (e.g., core $CC6$, PCIe $L1$, DRAM self-refresh), $PC1A$ leverages shallower power states (e.g., core $CC1$, PCIe $L0s$, DRAM CKE-off) and enables significant power savings with a nanosecond-scale exit latency.
Particularly, APMU orchestrates the $PC1A$ flow by interfacing with five key SoC components (as shown in \autoref{fig:pc1_arch}): 
1) CPU cores, 
2) high-speed IOs (PCIe, DMI, and UPI), 
3) memory controller,
4) CLM FIVR and clock tree, and  
5) global PMU (GPMU).

We place the APMU in north-cap, close to the firmware-based GPMU and IO domain~\cite{tam2018skylake,Skylake_die_server}. APMU implements three key power management infrastructure components.
First, a hardware fast (nanosecond granularity) \emph{finite-state-machine} (FSM) that orchestrates $PC1A$ entry and exit flows. The APMU FSM uses the same clock as the GPMU.

Second, \emph{status and event signals} that feed into the APMU FSM.
The {\tt InCC1} status signal combines (through AND gates) the status of all cores to notify the APMU that all cores are in the $CC1$ power state.
Similarly, the {\tt InL0s} status signal notifies the APMU that all IOs are in $L0s$ power state (see \autoref{sec:ASM}).
The GPMU {\tt WakeUp} signal sends a wakeup event to the APMU when an interrupt (e.g., timer expiration) occurs.
The {\tt PwrOk} signal notifies the APMU when the CLM FIVR reaches its target operational voltage level after exiting retention mode (see \autoref{sec:CCSM}).  

Third, \tech implements \emph{control signals} that the APMU uses to control \tech components.
The {\tt Allow\_CKE\_OFF} control signal, when set, enables the MC to enter {\it CKE off} low power state and to return to active state when unset. Similarly, the {\tt AllowL0s} signal, when set, enable the IO interfaces to enter $L0s$ power state and to return to active state when unset (see \autoref{sec:ASM}).
When {\tt Ret} signal is set, the CLM FIVRs reduce their voltage to pre-programmed retention level and they restore the previous voltage level when {\tt Ret} is unset (see \autoref{sec:CCSM}).
The APMU notifies the GPMU that the system in $PC1A$ by setting the {\tt InPC1A} signal.

\begin{figure}[h]
    \centering
	\includegraphics[trim=0.65cm 0.6cm 0.6cm 0.6cm, clip=true,width=0.95\linewidth,keepaspectratio]{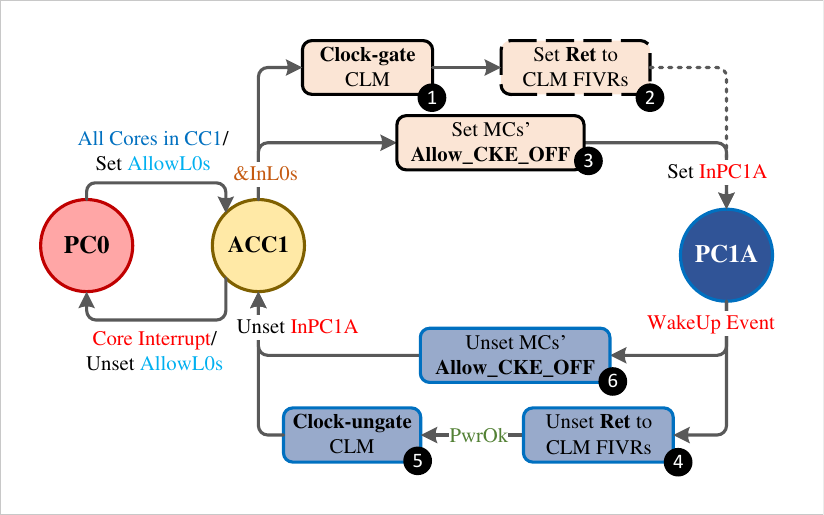}
	\vspace{-1pt}
	\caption{Power management flow for the $PC1A$ C-state.}
	\label{fig:pc1a_c_state_flows}
	\vspace{-1pt}
\end{figure}

\noindent{\bf PC1A Entry and Exit Flows.}
\tech power management flow, implemented by the APMU, is responsible for orchestrates the transitioning between $PC0$ and $PC1A$, as depcited in \autoref{fig:pc1a_c_state_flows}.
%

The $PC1A$ flow first transitions from $PC0$ to an intermediate state, $ACC1$, as soon as all cores enter $CC1$.
$ACC1$ serves the purpose of setting the {\tt AllowL0s} signal, thus allowing IOs to enter the $L0s$ state.
Next, once all IOs in $L0s$ state ({\tt \&InL0s}) the flow performs two branches, (i) and (ii),  concurrently:
(i) \circledb{1} it clock-gates the CLM and \circledb{2} initiates a non-blocking CLM voltage transition (by setting the {\tt Ret} signal) to reduce the voltage to retention level.
(ii) \circledb{3} it sets {\tt Allow\_CKE\_OFF} to allow the MCs to enter {\it CKE off}.
In contrast to existing package C-states (e.g., $PC6$, shown in \autoref{fig:pc6_c_state_flow}), the flow keeps all system phase-locked loops (PLLs) powered-on.
After these two steps (\circledb{2} is non-blocking) the system is in the $PC1A$ C-state.

Exiting $PC1A$ can happen because of two main causes:
First, an  IO link generates a \emph{wakeup event} when exiting $L0s$ due to traffic arrival; as soon as the link starts the transition from $L0s$ to $L0$, the IO link unsets {\tt InL0s}, generating a wakeup event in the APMU.
Second, the GPMU generates an explicit wakeup event by setting the {\tt WakeUp} signal.
The GPMU generates a wakeup event for multiple reasons, such as an interrupt, timer expiration, or thermal event.


When a \emph{wakeup event} occurs, the system \emph{exits} the $PC1A$ by reversing the entry flow in two branches, (i) and (ii),  concurrently:
(i) \circledb{4} it unsets the {\tt Ret} signal to ramp up the CLM voltage to its original level;
when the FIVRs set {\tt PwrOk}, \circledb{5} the flow clock-ungates the CLM. 
(ii) \circledb{6} it unsets {\tt Allow\_CKE\_OFF} to reactivate the MCs.
Once both branches are completed, the flow reaches the $ACC1$ state.  
Finally, in case the wakeup event is a \emph{core interrupt}, the interrupted core transitions from $CC1$ to $CC0$, correspondingly transitioning the system from $ACC1$ state to $PC0$ active state.
At this step, the flow unsets { \tt AllowL0s} to bring the IO links back to the active $L0$ state.

\vspace{-4pt}
\subsection{\asm~(\ASM)} 
\label{sec:ASM}

\ASM leverages IO shallow power states (e.g., $L0s$, CKE off) to enable significant power savings in $PC1A$ with a nanosecond-scale exit latency. We discuss PCIe, DMI, and UPI shallow power states in \autoref{sec:io_shallow_ps} and DRAM shallow power mode in \autoref{sec:dram_sr}

\vspace{-4pt}
\subsubsection{PCIe, DMI, and UPI in Shallow Power States}\label{sec:io_shallow_ps}

Once an IO interface is idle (i.e., not sending or receiving any transaction), the IO link and controller can enter to idle power state, called L-state, as explained in \autoref{sec:pm_states}.
Deep L-states ($L1$) have an exit latency of several $\mu$s, making them unsuitable for \tech.
Instead, we allow links to enter the $L0s$\footnote{PCIe, DMI, and QPI (UPI's previous generation) support $L0s$. UPI support $L0p$ state \cite{oneintel,lenovo_cstates,gough2015cpu}. We allow the IO to enter to $L0p$ in case it does not support $L0s$. $L0p$ exit latency is ${\sim}10ns$ \cite{gough2015cpu}.} state, which has exit latency in the order of tens of nanoseconds (e.g., $~64ns$).
While $L0s$ could be entered while other agents are active, datacenter servers normally completely disable it to avoid performance degradation~\cite{lenovo_cstates,dell_cstates,cisco_cstates}.
For the same reason, \tech keeps $L0s$ disabled when cores are active and allows high-speed IOs (e.g., PCIe, DMI, and UPI) to enter $L0s$ only when all the cores are idle (i.e., all cores in $CC1$). 

\noindent \textbf{AllowL0s Signal.}
To only allow entering $L0s$ when all cores are idle, \tech requires a new signal, {\tt AllowL0s} (light blue in \autoref{fig:pc1_arch}), to each IO controller.
The power management \emph{sets} the signal once all cores are in $CC1$ and each IO controller autonomously initiates the entry to $L0s$ state once the IO link is idle (i.e., no outstanding transaction)~\cite{gough2015cpu}. To allow the IO controller to enter quickly to $L0s$ once the IO link is idle, the  {\tt AllowL0s} signal also sets the \emph{L0s entry latency}\footnote{When $L0s$ is enabled and the link is in the $L0$ state, the IO controller transitions the link to the $L0s$ state if idle conditions are met for a period of time specified in the \emph{L0s Entry Latency} \cite{l0s_entry_lat}.} (L0S\_ENTRY\_LAT \cite{l0s_entry_lat}) configuration register.  Setting L0S\_ENTRY\_LAT to ``1'' sets the entry latency to $1/4$ of the $L0s$ exit latency, which is typically ${<}64ns$ \cite{l0s_entry_lat,icelake2020L0s}).    

\noindent \textbf{InL0s Indication.}
In the baseline system, the IO link power status (i.e., $L0$, $L0s$, and $L1$) is stored in a register inside the IO controller~\cite{pcie_ps}. Therefore, when entering a package C-state, the power management firmware needs to read this register. 
To make the new $PC1A$ agile, we add an output signal, {\tt InL0s} (orange in \autoref{fig:pc1_arch}), to each one of the high-speed IO controllers.
The IO link layer \emph{sets} the signal if the IO is at $L0s$ or deeper%
\footnote{In case that no device is present, the link power state is known as: \emph{no device attached} (NDA), which is deeper than $L1$ state~\cite{PCIE_NDA}.} %
and \emph{unsets} it if the link is in active state (i.e., $L0$) or is exiting the idle state.
The IO controller should \emph{unset} the signal once a \emph{wakeup} event is detected to allow the other system components to exit their idle state during $PC1A$ exit flow \emph{concurrently}; this transition only requires tens of nanoseconds.

\subsubsection{DRAM in a Shallow Power State}\label{sec:dram_sr}

When entering existing deep package C-states (e.g., $PC6$), the flow allows the memory controller to put DRAM into self-refresh mode (as shown in \autoref{fig:pc6_c_state_flow}).
The exit latency from self-refresh mode is several microseconds (see \autoref{sec:pm_states}) and unsuitable for $PC1A$.

\noindent \textbf{Allow\_CKE\_OFF Signal.}
Instead of using the long latency self-refresh mode, \tech instructs the memory controller (MC) to put DRAM into {\it CKE off} mode, which has lower power savings compared to self-refresh mode but massively lower exit latency (${<}30ns$).
To enable this transition, \tech adds a new input signal, {\tt Allow\_CKE\_OFF} to each memory controller (purple in \autoref{fig:pc1_arch}).
When this signal is set, the memory controller enters {\it CKE off} mode as soon as it completes all outstanding memory transactions and returns to the active state when unset.

\subsection{\ccsm~(\CCSM)}
\label{sec:CCSM}

In our reference, skylake-based multicore design, the last-level cache (LLC) is divided into multiple tiles, one per core, as \autoref{fig:skx_floorplan}(b) and \autoref{fig:pc1_arch} illustrate.
Each tile includes a portion of the LLC memory, a caching and home agent (CHA) and a snoop filter (SF); a mesh network-on-chip (NoC) connects the tiles with the IOs and memory controllers (MCs)~\cite{tam2018skylake}.
Two FIVR voltage domains (Vccclm0 and Vccclm1) power the CHA, LLC, and the (horizontal%
\footnote{The  Vccio (\autoref{fig:skx_floorplan}[c]) fixed voltage domain, delivered from a motherboard voltage regulator, powers the vertical mesh \cite{tam2018skylake}.}%
) mesh interconnect (known as CLM), as illustrated in \autoref{fig:skx_floorplan}(c).
When entering existing deep package C-states (i.e., $PC6$), the GPMU firmware turns off the phase-locked loop (PLL) for the CLM and reduces the Vccclm voltage to retention level to reduce leakage power. During $PC6$ exit, the firmware 1) send messages to the FIVRs to ramps up the Vccclm voltage and 2) re-locks the PLL (few microseconds).

To cut the time of re-locking the CLM PLL, \tech keeps the PLL locked and uses a new {\tt ClkGate} signal (brown in Fig.~\ref{fig:pc1_arch})  to allow quickly clock gating CLM's clock distribution network (e.g., clock tree).
To allow agile power management response, \tech adds a new signal, {\tt Ret} to each CLM FIVRs (green in Fig.~\ref{fig:pc1_arch}).
When {\tt Ret} is \emph{set}, the two CLM FIVRs reduce the voltage to pre-programmed retention voltage;
when {\tt Ret} is \emph{unset}, the FIVRs ramp their voltage back to the previous operational voltage level.
Once the FIVR voltage level reach the target, the FIVR \emph{sets} the {\tt PwrOK} signal.
\section{Implementation and HW Cost}
\label{sec:impl}

\tech requires the implementation of three main components: the \ASM subsystem, the \CCSM subsystem, and the agile power management unit (APMU).
We discuss implementation details for each component, including area and power cost, and the transition latency for the new $PC1A$ state.

\subsection{\asm~(\ASM)}\label{sec:imp_asm}

\ASM requires the implementation of \emph{three} signals depicted in \autoref{fig:pc1_arch}: 1) {\tt AllowL0s} (light blue), 2) {\tt InL0s} (orange), and 3) {\tt Allow\_CKE\_OFF} (purple). 

Implementing {\tt AllowL0s} requires routing control signals from the APMU to each one of the high-speed IO controllers (i.e., PCIe, DMI, and UPI). In each IO controller, the {\tt AllowL0s} control signal overrides the control register (e.g., LNKCON.active\_state\_link\_pm\_control \cite{pcie_l0s_reg})
 that prevents%
 \footnote{To maximize performance, vendors recommend to 
 disable IO power management and link state, including $L1$ and $L0s$ \cite{lenovo_cstates,dell_cstates,cisco_cstates}.}
 the \emph{Link Training and Status State Machine} (LTSSM)%
 \footnote{The LTSSM FSM manages the link operation of each high-speed IO \cite{chandana2015link,budruk2004pci,moreira2010engineer}.}
 from entering $L0s$ when the IO link is idle \cite{chandana2015link,budruk2004pci,moreira2010engineer}. We implement {\tt InL0s} using the LTSSM status: the IO controller sets {\tt InL0s} once the LTSSM reaches the $L0s$ state and unset it once the LTSSM exits the $L0s$ (i.e., a wakeup event is detected). The {\tt InL0s} output of each IO controller is routed to the APMU. To reduce routing overhead, the {\tt InL0s} of neighbouring IO controllers are aggregated using AND gates and routed to the APMU, as shown in \autoref{fig:pc1_arch}.  
 
 Similarly, implementing {\tt Allow\_CKE\_OFF}  requires routing a control signal from the APMU to each of the two memory controllers, as shown in \autoref{fig:pc1_arch}. The {\tt Allow\_CKE\_OFF} control signal overrides the control register in the memory controller (e.g., MC\_INIT\_STAT\_C.cke\_on \cite{pcie_l0s_reg})  that prevents an idle memory controller entering CKE off mode.
 
 Overall, \ASM adds \emph{five} long distance signals. In comparison to the number of data signals in an IO interconnect (mesh or ring), which typically has 128-bit -- 512-bit data width \cite{fallin2011high,alazemi2018routerless}, the additional five signals represent $1$\,--\,$4\%$ extra IO interconnect area. We extrapolate the IO interconnect area from a SKX die. The IO interconnect in north-cap~\cite{tam2018skylake}) is less than $6\%$ of SKX die area. Thus, the area overhead of the five new signals is \emph{${<}0.24\%$/${<}0.06\%$ of SKX die area} (assuming  128-bits/512-bits IO interconnect width).
This is a pessimistic estimate, since the IO interconnect includes control signals in addition to data.
 
Implementing the additional signals in the high-speed IOs (i.e., {\tt AllowL0s} and {\tt InL0s}) and the memory (i.e., {\tt Allow\_CKE\_OFF}) controllers only requires small modifications, since the required control/status knobs/signals are already present in the controllers.
Based on a comparable power-management flow implemented in~\cite{haj2020techniques}, we estimate the area required to implement the signals to be less than $0.5\%$ of each IO controller area. Given that the IO controllers take less than $15\%$ of the SKX die area, these signals will need \emph{less than $0.08\%$ of the SKX die area}.

\subsection{\ccsm~(\CCSM)}\label{sec:impl_ccsm}

Implementing \CCSM requires two main components 1) CLM clock-tree gating  and 2) CLM voltage control.
To allow clock gating/ungating of the CLM clock-tree, we route a control signal {\tt ClkGate} from the APMU to the existing CLM clock-tree control logic.
To control the CLM FIVRs voltage, we route an additional control signal, {\tt Ret}, from the APMU to the two FIVRs that power the CLM~\cite{tam2018skylake}.
To enable a FIVR to directly transition to a pre-programmed retention voltage, we add to each FIVR control module (FCM~\cite{nalamalpu2015broadwell,burton2014fivr}) an 8-bit \emph{register} that holds the retention voltage identification (RVID) value \cite{radhakrishnan2021power,luria2016dual}.
Finally, we add a {\tt PwrOk} status signal that the FIVR uses to notify the APMU that the voltage is stable.
%
Overall, \CCSM adds {three} long distance signals. Using analogous analysis as in \autoref{sec:imp_asm}, we estimate the area overhead for the three new signals is \emph{${<}0.14\%$ of SKX die area}. To implement the new RVID 8-bit register in each FIVR's FCM and add a new logic to select between the RVID and the original VID, needs less than $0.5\%$ of the FCMs' area. The FIVR area is less than $10\%$ of the SKX core die area and a core area in a die with 10 cores will be less than $10\%$ of the  SoC area, so the overall area overhead of two FCMs is negligible (less than 0.005\%).     


\vspace{-4pt}
\subsection{Agile Power Management Unit (APMU)} \label{sec:impl_flow}


The APMU, is implemented using a simple finite-state-machine (FSM) connected to the global PMU (GPMU), as depicted in \autoref{fig:pc1_arch}. APMU monitors its input status signals and drives its control signals as shown in \autoref{fig:pc1a_c_state_flows}.
Based on a comparable power-management flow implemented in 
~\mbox{\cite{haj2020techniques}}, we estimate the area required for the PC1A controller to be up to $5\%$ of the GPMU area. As shown in \autoref{fig:skx_floorplan} (dark blue), the GPMU area is less than $2\%$ of the SKX die area. Therefore, APMU area is \emph{less than $0.1\%$ of the SKX die area}.


We also need to implement a global status signal, {\tt InCC1}, that determines when all the CPU cores are at $CC1$ power state. The power state of each core is known to each core's power management agent (PMA~\mbox{\cite{rotem2012power}}), therefore, we simply expose this status as an output signal from each CPU core. The {\tt InCC1} output of each CPU core is routed to the APMU. To save routing resources, the {\tt InCC1} of neighbouring cores are combined with AND gates and routed to the APMU, as shown in blue in  \autoref{fig:pc1_arch}.
In total we we have $three$ long distance signals; according to our analysis in \autoref{sec:imp_asm}, their area overhead is ${<}0.14\%$ of the SKX die area. 

\textbf{In summary, the three \tech components discussed in Sections \ref{sec:imp_asm}, \ref{sec:impl_ccsm}, and \ref{sec:impl_flow} incur ${<}0.75\%$ overhead relative to a SKX die area.
}
\vspace{-4pt}
\subsection{PC1A Power Consumption Analysis}\label{sec:idle_power_analysis}

To estimate the $PC1A$ power, we carry out multiple measurements of our reference system (configuration in \autoref{sec:methodology}) to isolate the individual components contributing to the $PC1A$ power consumption.
As shown in \autoref{tbl:uarch_state_pcx}, the power consumption difference between  $PC1A$ and $PC6$ is due to the:   
1) CPU cores ($P_{cores\_diff}$), 
2) IOs power ($P_{IOs\_diff}$),
3) PLLs  ($P_{PLLs\_diff}$), and
4) DRAM ($P_{dram\_diff}$). 
Therefore, the $PC1A$ SoC power, $Psoc_{PC1A}$,  can be estimated as in \autoref{eq:pc1a-pow-soc}.

 \vspace{-10pt}
\begin{align}
  \resizebox{0.90\hsize}{!}{$Psoc_{PC1A} = Psoc_{PC6} + P_{cores\_diff}  + P_{IOs\_diff} + P_{PLLs\_diff}$} \label{eq:pc1a-pow-soc}
\end{align}
Similarly, the $PC1A$ DRAM power consumption, $Pdram_{PC1A}$, can be estimated as in \autoref{eq:pc1a-pow-dram}:

\vspace{-9pt}
\begin{align}
  \resizebox{0.56\hsize}{!}{$Pdram_{PC1A} = Pdram_{PC6} + P_{dram\_diff}$} \label{eq:pc1a-pow-dram}
\end{align}

We use Intel's RAPL monitoring interface~\cite{khan2018rapl,fahad2019comparative,hackenberg2015energy} to measure the SoC (package) and DRAM power consumption. Next, we discuss the \emph{two} configurations we use to determine each one of the four power deltas between $PC1A$ and $PC6$.  

\noindent $\mathbf{P_{cores\_diff}}$: To measure the cores power difference between our new $PC1A$ and $PC6$, denoted by $P_{cores\_diff}$, we use two system configurations: 1) all cores are placed in in $CC1$  and 2) all cores are placed in in $CC6$. To keep  uncore power consumption similar in the two configurations, we disable uncore power savings techniques such as package $C6$, DRAM opportunistic self-refresh (OSR) , memory power-down (CKE off), uncore frequency scaling~\cite{gough2015cpu,osr_self_refresh,lenovo_cstates}. We measure the power of the two configurations using RAPL.Package~\cite{khan2018rapl,fahad2019comparative,hackenberg2015energy}\footnote{An alternative measuring option is to use RAPL.PP0, which is the aggregate total of all cores. However, RAPL.PP0 is not available in our system, a known issue reported by users~\cite{rapl_pp0_not_working1,rapl_pp0_not_working2}.} and calculate the difference. 
Our measurements shows that $P_{cores\_diff}\approx 12.1W$

\noindent {\bf $\mathbf{P_{IOs\_diff}}$ and $\mathbf{P_{dram\_diff}}$}: The IOs power includes PCIe, DMI, UPI, and memory controllers and their corresponding physical layers (PHYs) but it does not include the devices' (e.g., DRAM) power. To measure the IOs power consumption difference between $PC1A$ and $PC6$, denoted by $P_{IOs\_diff}$, we use two configurations: 1) place the  PCIe and DMI in  $L0s$ power state, UPI to $L0p$ power mode, and memory-controller (MC) in $CKE off$ power mode and 2) place the PCIe, DMI, and UPI in $L1$ power state, and memory-controller (MC) in self-refresh power mode. To place the system in these power modes, we use BIOS configurations to  i) place the cores in core $CC6$ and set the package C-state limit  to PC2 to allow the IOs to enter to local power mode but prevent the system from entering $PC6$~\cite{ISB_bios_guide}, ii) set the PCIe/DMI/UPI active state power management to $L0s$/$L0s$/$L0p$ for the first configuration and to $L1$/$L1$/$L1$ for the second configuration~\cite{lenovo_cstates}, and iii) configure the memory to enter power-down (CKE off) and opportunistic self refresh (OSR)~\cite{osr_self_refresh,gough2015cpu,lenovo_cstates} for the first and second configuration, respectively. To obtain $P_{IOs\_diff}$ ($P_{dram\_diff}$) we measure the power of the two configurations using RAPL.Package (RAPL.DRAM)~\cite{khan2018rapl,fahad2019comparative,hackenberg2015energy} and calculate the difference.
Our measurements shows that $P_{IOs\_diff}\approx 3.5W$ and $P_{dram\_diff}\approx 1.1W$

\noindent $\mathbf{P_{PLLs\_diff}}$: All PLLs are on in $PC1A$, but off in $PC6$. We estimate the PLLs power consumption difference between our new $PC1A$ and $PC6$, denoted by $P_{PLLs\_diff}$, by: \emph{number of system PLLs} times \emph{a PLL power}. In our SKX system~\cite{Skylake_4114} there are approximately $18$ PLLs: one PLL for each PCIe, DMI, and UPI controller~\cite{intel_2nd_gen_xeon_datatsheet} (our system~\cite{Skylake_4114} has $3$ PCIe, $1$ DMI, and $2$ UPI), one PLL for the CLM and memory controllers~\cite{tam2018skylake}, one PLL for the global power management unit \cite{Skylake_die_server}, and one PLL per core ($10$ cores in our system~\cite{Skylake_4114}). The per core PLL power is accounted for in $P_{cores\_diff}$, since we measure RAPL.Package. Therefore, there are $8$ remaining PLLs. 
The Skylake system uses all-digital phase-locked loop (ADPLLs)~\cite{fayneh20164,Skylake_die_server} that consume $7$mW each (fixed across core voltage/frequency~\cite{fayneh20164}). Therefore, the estimated  $P_{PLLs\_diff}$ power is $56mW$.

 We place the system in $PC6$ state and using RAPL.Package and RAPL.DRAM we measure $Psoc_{PC6}$ ($11.9W$) and $Pdram_{PC6}$ ($0.51W$), respectively. In summary, $Psoc_{PC1A}\approx 11.9W + 12.1W + 3.5W +0.057W \approx 27.5W$  and $Pdram_{PC1A}\approx 0.51W + 1.1W\approx 1.6W$, as we summarize in \autoref{tab:c-states}.


\vspace{-4pt}
\subsection{PC1A Latency}\label{sec:c6a_lat}

We estimate that the overall transition time (i.e., entry followed by direct exit) for the \tech's $PC1A$ state  to be ${<}200$ns: ${>}250\times$  faster than the ${>}50${\textmu}s that $PC6$ requires.
Next, we discuss in detail the entry and exit for $PC1A$; we refer to the power management flow shown in \autoref{fig:pc1a_c_state_flows}.

\vspace{-5pt}
\subsubsection{PC1A Entry Latency} 
The package C-state flow starts once all cores are idle; when all the cores enter to $CC1$, the system transitions to $ACC1$ package state. Similar to the traditional $PC2$ package C-state (shown in \autoref{fig:pc6_c_state_flow}, $ACC1$ is a temporary state at which uncore resources (LLC, DRAM, IOs) are still available. Therefore, we measure $PC1A$ latency starting from the $ACC1$. 

In $ACC1$, we enable the IOs to enter a shallow power state (i.e., $L0s$). As discussed in \autoref{sec:io_shallow_ps}, the entry latency of the IO (PICe, DMI, and UPI) controllers is $\approx25\%$ of the exit latency (typically ${<}64ns$). Therefore, once the IOs are idle for $16ns$ the IO enters $L0s$ state and sets the \texttt{InL0s} signal. In case some IOs are not idle, the system remains in $ACC1$. When an interrupt occurs, the system moves back to $PC0$.  

Clock-gating the CLM domain and keeping the PLL ON \circledb{1} typically takes $1$\,--\,$2$ cycles in an optimized clock distribution system~\cite{el2011clocking,shamanna2010scalable}.
Reducing CLM's voltage \circledb{2} from nominal voltage (${\sim}0.8V$) to retention voltage (${\sim}0.5V$)~\cite{chen201322nm,chen201322nm_ppt}, is a non-blocking process. FIVR's voltage slew rate is typically typically ${\geq}{2mV}/ns$~\cite{burton2014fivr,kar2017all}. Thus, the time it takes for the FIVR to reduce the voltage by $300mV$ (from ${\sim}0.8V$ to ${\sim}0.5V$) is ${\leq}150ns$. Asserting  MCs' $Allow\_CKE\_OFF$ control signal takes $1$\,--\,$2$ cycles. Since the system is idle, once  the MCs receive the $Allow\_CKE\_OFF$ signal they enter CKE off within $10ns$~\cite{david2011memory,malladi2012rethinking}.            

In summary, since voltage transition to retention and entry to CKE off mode are non-blocking, $PC1A$ entry latency is ${\sim}18ns$ using a power management controller with $500MHz$ clock frequency.\footnote{Power management controllers of a modern SoCs operate at clock frequency of several megahertz (e.g., 500MHz \cite{peterson2019fully}) to handle nanosecond-scale events, such as \textit{di/dt}  prevention \hp{\mbox{\cite{fayneh20164}\cite[Sec. 5]{haj2021ichannels}}}.}

\vspace{-5pt}
\subsubsection{PC1A Exit Latency} 

$PC1A$ exit is caused by wakeup events (e.g., IO activity, GPMU timer).
In case of IO events, the IO links concurrently start exiting $L0s$/$L0p$ (a process that requires ${<}64ns$) and a wake-up event is signaled to the APMU.  

Increasing the CLM's voltage \circledb{4} from retention (${\sim}0.5V$) to nominal voltage (${\sim}0.8V$)~\cite{chen201322nm,chen201322nm_ppt}, takes $150ns$ since FIVR's voltage slew rate is typically ${\geq}{2mV}/ns$ \cite{burton2014fivr,kar2017all}.\footnote{We assume FIVR with preemptive voltage commands: to allow fast C-state exit latency, a modern VR implements preemptive voltage commands; In which the VR interrupts its current voltage transition to first $VID_1$ and moves to handle a new request to second $VID_2$ (e.g., once a C-state entry flow is interrupted and the flow need to exit in the middle of a voltage transition to retention)~\cite{oneintel}.}  
Clock-ungating the CLM domain and keeping the PLL ON \circledb{5} typically takes $1$\,--\,$2$ cycles in an optimized clock distribution system~\cite{el2011clocking,shamanna2010scalable}.
Unsettting  MCs' \texttt{Allow\_CKE\_OFF} control signal \circledb{6} takes $1$\,--\,$2$ cycles. Once the MCs receive the \texttt{Allow\_CKE\_OFF} signal, they exit MCs CKE off mode within  $24ns$ \cite{david2011memory,malladi2012rethinking,appuswamy2015scaling}.

In summary,  $PC1A$ exit latency is ${\leq}150ns$ using a power management controller with $500MHz$ clock frequency. 
The worst case entry plus exit latency is ${\leq}168ns$. We conservatively assume ${\leq}200ns$.

\vspace{-4pt}
\subsection{Design Effort and Complexity}
\tech proposed techniques involve non-negligible front-end and back-end design complexity and effort. 
The APMU,  $PC1A$ \emph{control flows}, \ASM, and \CCSM, require careful pre-silicon  verification to ensure that all the hardware flows (described in \autoref{fig:pc1a_c_state_flows}), IO controllers (PCIe, DMI, UPI, MC), and CPU core changes operating as expected by the architecture specification. The effort and complexity can be significant due to two main reasons. 
1) \tech involves system-on-chip global changes, requiring  careful coordination between multiple design teams.
2) the power management flows are hardware-based, which, compared to firmware-based flow, reduces the opportunity to patch the flows if a hardware bug is found post-silicon production.    

However, \tech effort and complexity are comparable to recent techniques implemented in modern processors to increase their energy efficiency (e.g., hybrid cores \cite{rotem2021alder,apple_m1}). Therefore, we believe that once there is a strong request from customers and/or pressure from competitors, several processor vendors will eventually implement a similar architecture to \tech to  increase server energy efficiency significantly.

\vspace{-4pt}
\section{Experimental Methodology}
\label{sec:methodology}

We evaluate \tech using three latency-critical 
\dbb{services}: \emph{Memcached}, \emph{Apache Kafka}, and \emph{MySQL}.
Memcached~\cite{memcached} is a \dbb{popular} key-value store \dbb{commonly} deployed as a distributed caching \dbb{layer} to accelerate user-facing applications
~\cite{nishtala2013memcached, yang2020twemcache, pymemcache}.
\dbb{
Memcached has been widely studied~\cite{lim2013memcached, xu2014memcached, leverich2014mutilate, prekas2017zygos}, particularly for tail latency performance optimization~\cite{nishtala2013memcached, jialin2014memcached, asyabi2020peafowl}.}
Kafka~\cite{kreps2011kafka} is a real-time event streaming platform used to power event-driven microservices and stream processing applications.
MySQL~\cite{mysql} is a widely used relational database management system.

We use a small cluster of servers to run our three \dbb{services} and the corresponding clients.
Each server has an Intel Xeon Silver 4114~\cite{Skylake_4114} processor running at $2.2$\,GHz nominal frequency (minimum $0.8$\,GHz, maximum Turbo Boost frequency $3$\,GHz) with 10 physical cores (total of 20 hyper-threads) and $192$\,GB \dbb{of ECC} DDR4 \dbb{2666MHz} DRAM.

\noindent {\bf Workload setup.}
\dbb{For each of our three services (Memcached, Kafka, MySQL), we run a single server process on a dedicated machine and corresponding clients on separate machines.
We pin server processes to specific cores to minimize the impact of the OS scheduler.
The Memcached client is a modified version of the Mutilate load generator~\cite{leverich2014mutilate} set
to reproduce the ETC Facebook workload~\cite{atikoglu2012workload} using one master and four workload-generator clients, each running on a separate machine. 
The Kafka client consists of the ConsumerPerformance and ProducerPerformance Kafka. 
The MySQL client consists of the \texttt{sysbench} benchmarking tool using the OLTP test profile~\cite{sysbench}.} 

\ys{\noindent {\bf Baseline configurations.}
We consider two baseline configurations: $C_{shallow}$ and $C_{deep}$.
The $C_{shallow}$ configuration is representative of real modern datacenters that, as discussed in \autoref{sec:intro}, are normally configured for maximum performance~\cite{cisco_cstates,dell_cstates,lenovo_cstates}.
Therefore, in the $C_{shallow}$ configuration, we disable the CC6 and CC1E core C-states and all package C-states.
Additionally, we disable P-states (i.e., DVFS) by setting the frequency scaling governor to {\it performance} mode (i.e., nominal frequency), to avoid frequency fluctuations. 
The $C_{deep}$ configuration has all core and package C-states enabled.
P-states are still disabled, but the frequency scaling governor is set to {\it powersave} mode.
In order to allow the system to enter PC6, we tune it using the auto-tune option from {\it powertop}~\cite{powertop}.
We obtain C-state residency and number of transitions using residency reporting counters~\cite{intel_skl_dev}, and we use the RAPL interface~\cite{HDC_intel} to measure power consumption. }

\dbb{\noindent {\bf Power and performance models.}
We estimate the impact of the \tech on power and performance with a combination of simple models and real measurements.
We base power estimations on the same model as in  \autoref{eq:savings} (\autoref{sec:motivation}).
For the performance model, we calculate the impact on average latency by combining the number of $PC1A$ transitions, measured on our baseline system, with the additional transition latency required for $PC1A$ (see \autoref{sec:c6a_lat}).}


\dbb{
\noindent {\bf Power event tracing.}
We estimate the opportunity for PC1A residency using Intel's SoCWatch~\cite{socwatch} energy analysis collection tool.
We use SoCWatch to generate a trace that records C-state transition events, and we process this timeline to identify opportunities to enter $PC1A$. 
}
\dbb{Due to sampling constraints, SoCwatch does not \dbb{record} idle periods shorter than $10$\,us; therefore, the $PC1A$ opportunity we present in \autoref{sec:evaluation} underestimates the real opportunity.
%
%
We additionally use SoCWatch to measure the distribution of the number of active cores after full idle periods (i.e., periods during which all cores are in $CC1$ or lower C-state).
We use this metric and the $PC1A$ transitions to estimate the performance impact presented in \autoref{sec:evaluation}.}

\vspace{-4pt}
\section{Evaluation}
\label{sec:evaluation}

Our evaluation of \tech addresses the following questions:
%
\begin{enumerate}
\item What is the opportunity to enter \tech's new agile deep package C-state ($PC1A$)?
\item What are the power savings $PC1A$ can enable?
\item How does $PC1A$ impact performance?
\end{enumerate}
\dbb{We first focus on the Memcached~\cite{memcached} service and later discuss results on our two other workloads in \autoref{sec:other_workloads}.
We tune the client to generate a wide range of request intensity, but focus on the lower end (approximately $5\,-\,20\%$ processor utilization), which represents the typical operating range of servers running latency-critical applications~\cite{lo2014towards,B1,B2,B3,B4,iorgulescu2018perfiso}.
For Memcached, this load range corresponds to a range of $4K-100K$\,QPS (queries per second).
In our plots, we highlight the low-load region with a \emph{shaded area}.
}

\begin{figure}[t!]
\centering
\includegraphics[trim=0.1cm 0cm 0.1cm 0.1cm, clip=true,width=0.99\linewidth,keepaspectratio]{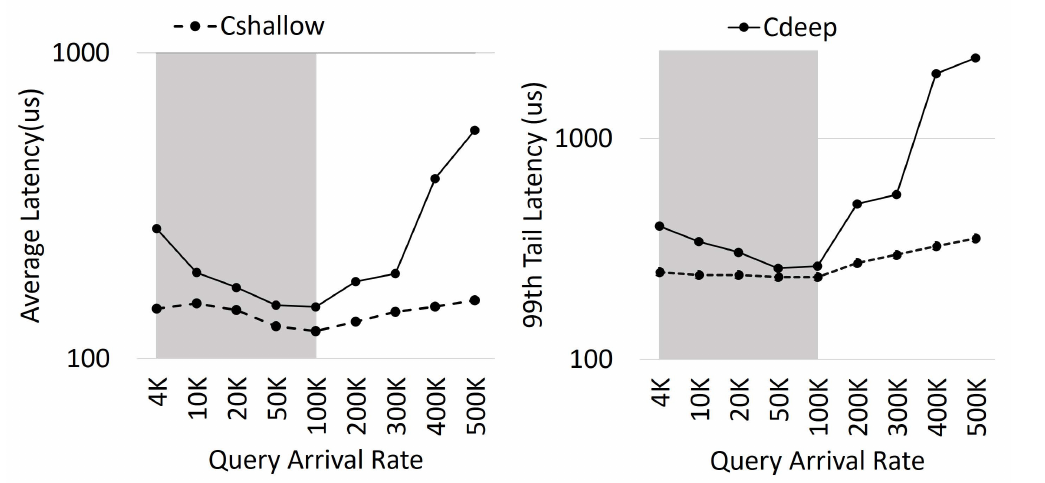}
\vspace{-8pt}
\caption{%
\dbb{Impact on Memcached from enabling deep C-states ($C_{shallow}$ vs $C_{deep}$) on average and tail latency.}}
\vspace{-10pt}
\label{fig:res_cc1e_cc6}
\end{figure}

\dbb{
First, we analyze the performance impact of enabling deep core C-states and package C-states on Memcached.
\autoref{fig:res_cc1e_cc6} compares the average and tail latency of Memcached running on the {\it $C_{shallow}$} configuration against the {\it $C_{deep}$} configuration (which enables deep C-states, see \autoref{sec:methodology}).

The $C_{shallow}$ configuration has significantly better average and tail latency compared to the $C_{deep}$ configuration}, as it avoids deep core C-state transition overhead, thus corroborating the advise of server manufacturers. 
However, the $C_{shallow}$ configuration also prevents entering any power-saving package C-state, thus missing the opportunity to save package power during periods of full system idleness. 
\hv{At high load ($\ge300K$ QPS) of the $C_{deep}$ configuration, we observe a latency spike caused by $CC6$/$PC6$ transitions delaying the processing of the initial incoming requests, which further delays and queues  following requests.
}
\dbb{
}

\vspace{-4pt}
\subsection{PC1A Opportunity}
\label{sec:pc1a-opportunity}

\begin{figure}[t!]
\centering
\includegraphics[trim=0.1cm 0cm 0cm 0cm, clip=true,width=0.99\linewidth,keepaspectratio]{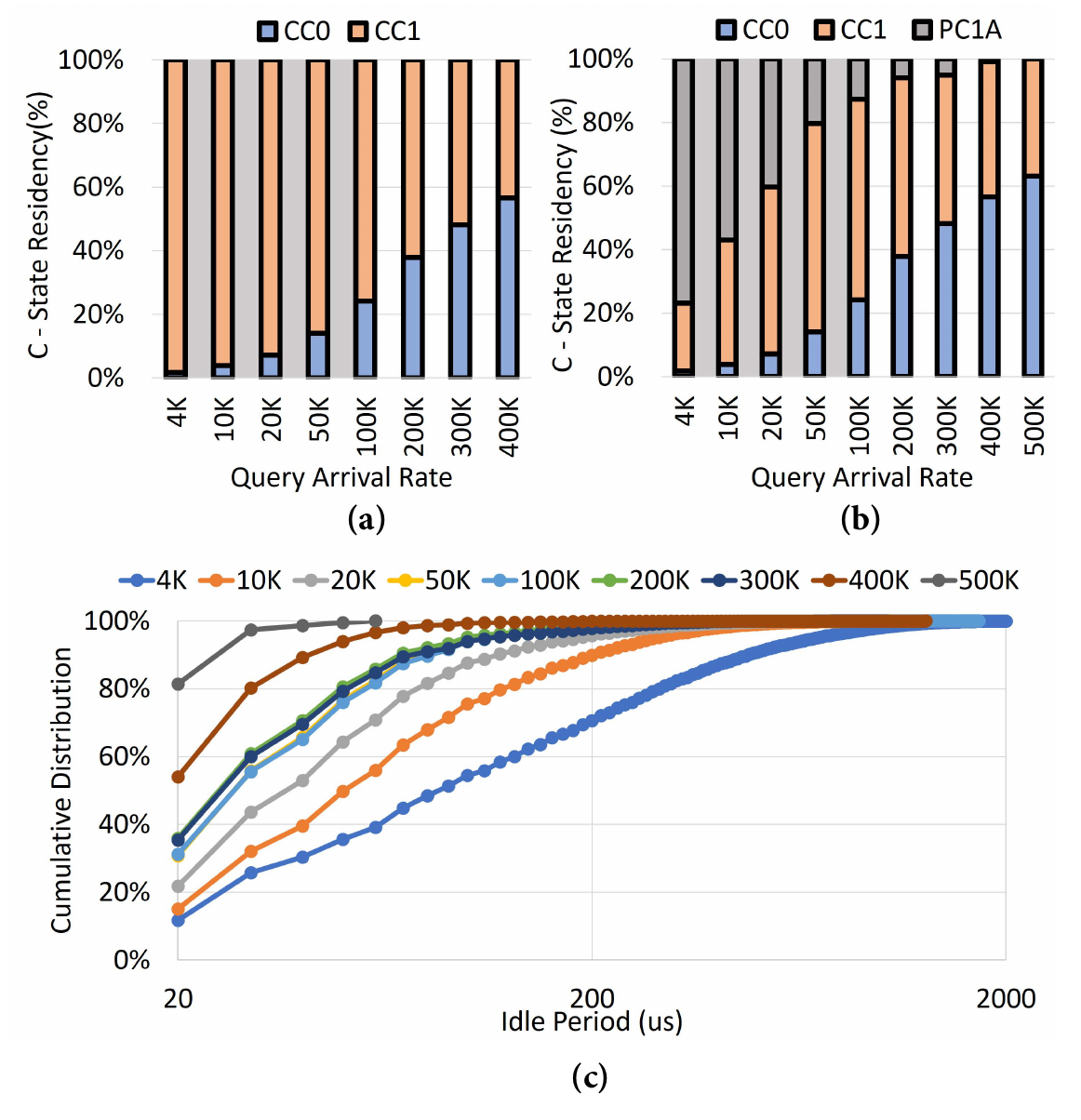}
\vspace{-10pt}
\caption{%
\rG{PC1A opportunity for Memcached. (a) Core C-state residency of the $C_{shallow}$ baseline. (b) PC1A residency (PC1A is fraction of time that all cores are in CC1). (c) Distribution of fully idle periods.}}
\vspace{-10pt}
\label{fig:res_cc1}
\end{figure}

\begin{figure*}[t]
\centering
\includegraphics[trim=0cm 0cm 0cm 0cm, clip=true,width=0.95\linewidth,keepaspectratio]{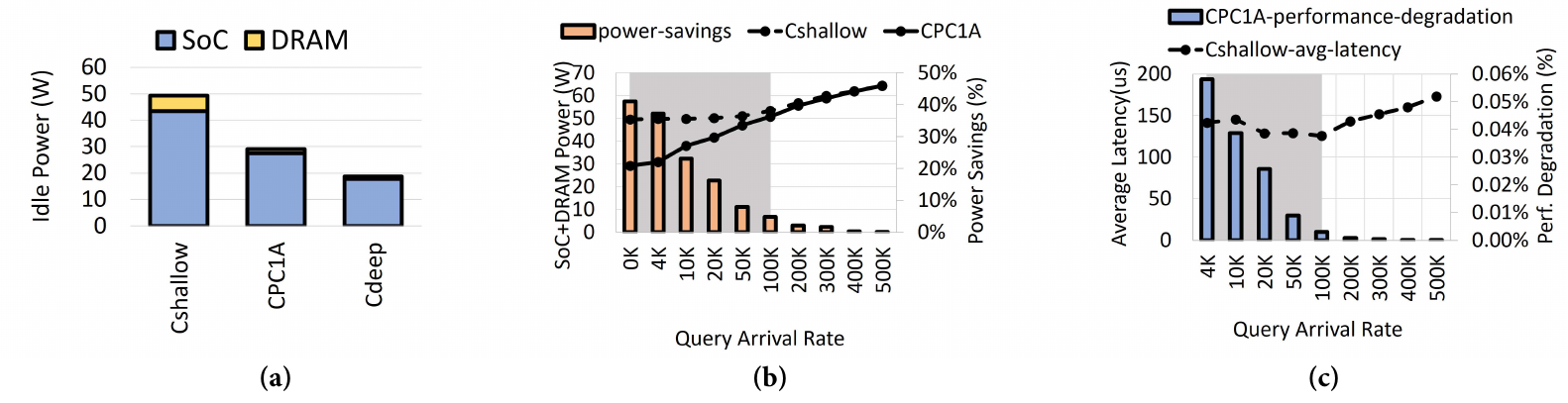}
\vspace{-10pt}
\caption{%
\rG{Power savings and performance impact from entering \tech's PC1A in Memcached. (a) Idle power consumption. (b) Power consumption and savings at different request rates. (c) Average latency and impact at different request rates.}}
\vspace{-12pt}
\label{fig:res_pc1a_power}
\label{fig:res_pc1a_latency}
\end{figure*}

\dbb{\autoref{fig:res_cc1} quantifies the opportunity for the system to enter \tech's new $PC1A$ package C-state as soon as all cores are in $CC1$.}
\autoref{fig:res_cc1}(a) shows core C-state residency for the $C_{shallow}$ baseline; \ys{the average fraction of time each core is in CC0 and CC1 core C-states}.
\ys{For low load ($\leq100K$ QPS), we observe that for a large fraction of time (at least $76\%$ to $98\%$) a core is in CC1.}
Entering PC1A, however, requires all cores to concurrently \hv{be present at CC1}; 
\autoref{fig:res_cc1}(b) quantifies this opportunity.
Since the baseline system, we use to emulate \tech, does not actually implement the PC1A state, we estimate PC1A residency 
as the fraction of time when the system is fully idle, i.e., all cores are simultaneously in $CC1$.
We collect this information through SoCwatch, as described in \autoref{sec:methodology}.
We observe that, although $PC1A$ residency diminishes at high load, the opportunity is significant ($\geq12\%$) at low load ($\leq100$ QPS), with
\dbb{PC1A residency reaching $77\%$ at $4$k QPS and $20\%$ for $50$k QPS.}
%
\autoref{fig:res_cc1}(c) provides further details on the distribution of the length \dbb{of fully idle periods (i.e., all cores in CC1)}.
We observe that, at low load, $60\%$ of the idle periods have a duration between $20\mu$s and $200\mu$s, 
\hj{whereas} \ys{the $PC1A$ transition latency is $\leq 200ns$. The fast $PC1A$ transition latency enables to reap most of the power reduction opportunity during short periods with all cores idle. This is infeasible with existing $PC6$ state, which has almost no power saving opportunity with its ${>}50\mu$s transition latency.}


Since servers running latency-critical applications typically operate at low load, we conclude that real deployments have significant opportunity to enter \dbb{\tech's new $PC1A$ C-state and benefit from its power savings, which we discuss next}.

\vspace{-4pt}
\subsection{PC1A Power Savings}

Having confirmed the opportunity to enter package C-state $PC1A$, we now study the power savings we can expect from \tech.
%
\autoref{fig:res_pc1a_power}(a) shows the processor SoC and DRAM power consumption when all cores are idle for three different configurations: 
$C_{shallow}$ baseline, 
$C_{deep}$ baseline, 
\hj{and \emph{C$_{PC1A}$}}. \ys{$C_{PC1A}$ corresponds to the $C_{shallow}$ configuration enhanced with our new $PC1A$ package C-state.}
We estimate idle package power and idle DRAM power of \ys{C$_{PC1A}$} using our power analysis \hj{discussed in} \autoref{sec:impl}.
\dbb{Idle power for the} \hj{C$_{PC1A}$} configuration \dbb{is at a middle point between the $C_{shallow}$ (i.e., no package power savings) and the $C_{deep}$ (i.e., deep C-states enabled, but unrealistic for servers).}
\dbb{More specifically, \ys{C$_{PC1A}$} enables} $41\%$ lower idle power consumption than the \dbb{$C_{shallow}$}.

\autoref{fig:res_pc1a_power}(b) reports \hj{1) the $C_{shallow}$ baseline and \ys{C$_{PC1A}$} power consumption, and 2) \ys{C$_{PC1A}$}'s} power savings \hj{as compared to \dbb{the $C_{shallow}$} baseline} \hj{for varying request rates (QPS)}. We observe that \ys{C$_{PC1A}$} has lower (or equal) power consumption than the baseline system across the entire range of request rates. The power savings are more pronounced at low load, where the opportunity to enter the $PC1A$ state is higher, as discussed in \autoref{sec:pc1a-opportunity}.
At $4$k QPS, the \ys{C$_{PC1A}$} configuration has $37\%$ lower power,
while at $50$K QPS, it has $14\%$ lower power.
The 0K QPS represents the \hj{expected} power savings during \hj{idle} periods\dbb{, when} \hj{no tasks are assigned to the server}. 

We conclude that the new deep package C-state, $PC1A$, results in significant power savings \dbb{during fully idle periods and at} low load, \hj{the operating points in which modern server have poor energy efficiency~\cite{lo2014towards}}, thus making the \hj{datacenter servers} more \emph{energy proportional}. 

\vspace{-4pt}
\subsection{PC1A Performance Impact}


Although 
PC1A makes the system more energy proportional, entering and exiting PC1A introduces a small \hj{(${<}200ns$)} transition \hj{overhead}. 
\autoref{fig:res_pc1a_latency}(c) analyzes the impact of \tech on \hj{\emph{average end-to-end  latency}} \hj{for different request rates}, according to \hj{our} methodology described in \autoref{sec:methodology}. 
End-to-end latency includes server-side latency plus network latency, \dbb{which accounts to $\approx117\mu$s}.

\hj{To estimate the performance degradation for different request rates, our performance model uses} \hj{1)} the number of $PC1A$ transitions\hj{, 2)} the distribution of number of active cores after \hj{exiting} full idle,  and \hj{3)} the transition cost ($200ns$).
We observe that even in the worst case, $PC1A$ has a \emph{negligible} impact ($< 0.1\%$) on average  latency.
\dbb{While we do not show additional results due to space constraints, we observe that } the \dbb{overhead on} \emph{\hj{end-to-end} tail  latency} \dbb{is} even smaller.  


\hj{We} conclude that $PC1A$ is a practical package C-state \hj{that improves energy proportionality \dbb{for datacenter servers}  with}  \emph{negligible} performance degradation. 

\vspace{-4pt}
\subsection{Analysis of Additional Workloads}
\label{sec:other_workloads}
\vspace{-8pt}

\begin{figure}[h]
\centering
\includegraphics[trim=0.1cm 0.1cm 0.1cm 0.1cm, clip=true,width=0.99\linewidth,keepaspectratio]{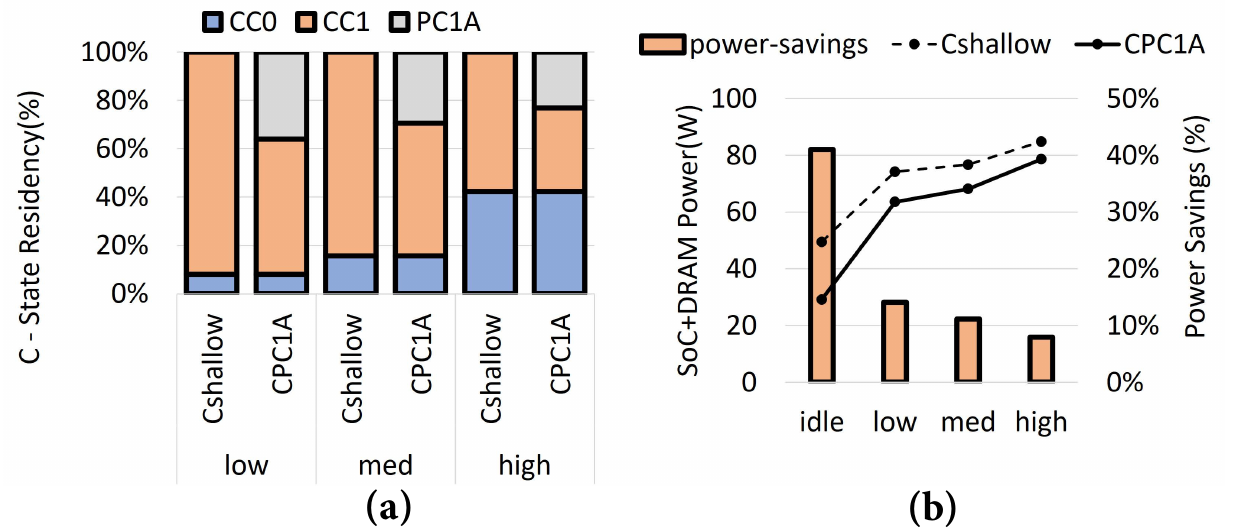}
\vspace{-4pt}
\caption{Evaluation of MySQL for low, mid and high request rates. 
(a) Residency of the \dbb{$C_{shallow}$} baseline and $C_{PC1A}$ at different core C-states and PC1A.
(b) Average power reduction of the $C_{PC1A}$ configuration as compared to the \dbb{$C_{shallow}$}.
}
 \vspace{-5pt}
\label{fig:res_mysql}
\end{figure}

\autoref{fig:res_mysql} shows the evaluation of MySQL \cite{mysql} for three request rates (low, mid, and high), corresponding to \ga{ $8\%$, $16\%$, and $42\%$} processor load.
\autoref{fig:res_mysql}(a) shows the \hj{core} C-state and projected PC1A residency of the \dbb{$C_{shallow}$} baseline and $C_{PC1A}$.
 We observe a notable opportunity to enter PC1A across all request rates. 
\dbb{The $C_{shallow}$ baseline spends \ga{$20\%$ to $37\%$} of the time with all cores idle (i.e., in $CC1$), translating in corresponding opportunity for $PC1A$ residency for $C_{PC1A}$.}
\autoref{fig:res_mysql}(b) \dbb{translates $PC1A$ residency to power savings, amounting to} \ga{$7\%$ to $14\%$} average power reduction with $C_{PC1A}$.

\begin{figure}[t]
\centering
\includegraphics[trim=0.1cm 0cm 0.1cm 0cm, clip=true,width=0.99\linewidth,keepaspectratio]{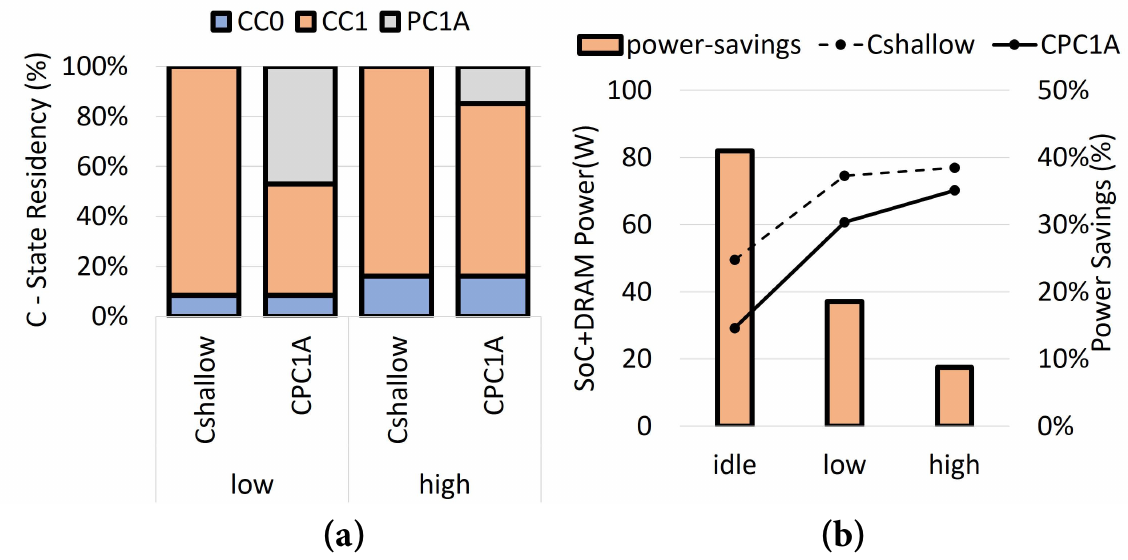}
 \vspace{-5pt}
\caption{Evaluation of Kafka for low and high request rates.
(a) Residency of the \dbb{$C_{shallow}$} baseline and $C_{PC1A}$ at different core C-states and PC1A.
(b) Average power reduction of the $C_{PC1A}$ configuration as compared to the \dbb{$C_{shallow}$}.
}
\vspace{-8pt}
\label{fig:res_kafka}
\end{figure}

\autoref{fig:res_kafka} \dbb{presents a similar analysis for} Kafka~\cite{kreps2011kafka} for two request rates (low and high), corresponding to \ga{$8\%$ and $16\%$} processor load.
%
\dbb{\autoref{fig:res_kafka}(a) shows opportunity to enter PC1A at both all load levels, reaching an estimated \ga{$15\%$ to $47\%$} $PC1A$ residency.
Fig. \ref{fig:res_kafka}(b) shows that the PC1A residency translates to \ga{$9\%$ to $19\%$} average power reduction from entering $PC1A$.
}

 When the server is fully idle, i.e., no tasks are assigned to the server, the average power reduction with $C_{PC1A}$ is $41\%$, as shown in \autoref{fig:res_mysql}(b) \autoref{fig:res_kafka}(b).
\dbb{We additionally analyze the performance impact and found that }
the impact of \tech on average and tail latency for both Kafka and MySQL is negligible (${<}0.01\%$).
\vspace{-4pt}
\section{Related Work}
\label{sec:related}

To our knowledge, \tech is the first practical proposal for a new package C-state design directly targeting latency-critical applications in datacenters.
While the problem of low server efficiency for latency-critical workloads has been studied before, previous work proposes management and scheduling techniques to mitigate the problem, rather than addressing it directly.
A low-latency {\it package} power-saving state is of key importance, since it not only enables power savings in uncore components in the SoC, but also in the whole system.

\noindent\textbf{Fine-grained, Latency-Aware DVFS Management.}
Besides C-states, the other major power-management feature of modern processors is dynamic voltage and frequency scaling (DVFS).
Previous work proposes fine-grained DVFS control to save power, while avoiding excessive latency degradation.
Rubik~\cite{kasture2015rubik} scales core frequency at sub-ms scale based on a statistical performance model to save power, while still meeting target tail latency requirements.
Swan~\cite{zhou2020swan} extends this idea to computational sprinting (e.g., Intel Turbo Boost): requests are initially served on a core operating at low frequency and, depending on the  load, Swan scales the frequency up (including sprinting levels) to catch up and meet latency requirements.
NMAP~\cite{kang2021nmap}, focuses on the network stack and leverages transitions between polling and interrupt mode as a signal to drive DVFS management.
The new \CAgile state of \tech facilitates the effective use of idle states and makes a simple race-to-halt approach more attractive compared to complex DVFS management techniques.

\noindent\textbf{Workload-Aware Idle State Management.}
Various proposals exist for techniques that profile incoming request streams and use that information to improve power management decisions.
SleepScale~\cite{liu2014sleepscale} is a runtime power management tool that selects the most efficient C-state and DVFS setting for a given QoS constraint based on workload profiling information.
WASP~\cite{yao2017wasp} proposes a two-level power management framework;
the first level tries to steer bursty request streams to a subset of servers, such that other machines can leverage deeper, longer-latency idle states;
the second level adjusts local power management decisions based on workload characteristics such as job size, arrival pattern and system utilization.
Similarly, CARB~\cite{zhan2016carb} tries to pack requests into a small subset of cores, while limiting latency degradation, so that the other cores have longer quiet times and can transition to deeper C-states.
The idea of packing requests onto a subset of active cores, so as to extend quiet periods on other cores is further explored by other work focusing on both C-state and DVFS management~\cite{chou2016dynsleep,asyabi2020peafowl,chou2019mudpm}.
These proposals are orthogonal to \tech and can bring additive improvements.
In particular, a technique that synchronizes active / idle periods across different cores while curbing latency degradation can increase the duration of system-level idle periods and, subsequently, the power-saving opportunity.


  \vspace{-4pt}
\section{Conclusion}

This paper presents the design of \techL(\tech): a new C-state architecture that improves the energy proportionality of servers that operate at low utilization while running microservices of user-facing applications. \tech targets the reduction of power when all cores are idle in a shallow C-state ready to transition back to service. In particular, \tech targets the power of the resources shared by the cores (e.g., LLC, network-on-chip, IOs, DRAM) which remain active while no core is active to use them. \tech realizes its objective by using low-overhead hardware to facilitate sub-microsecond entry/exit latency to a new package C-state and judiciously selecting intermediate power modes, for the different shared resources, that offer fast transition and, yet, substantial power savings. Our experimental evaluation supports that \tech holds potential to reduce server power of up to $41\%$ with a worst case performance degradation less than 0.1\% for several representative workloads. Our results clearly support for the research and development and eventual adoption of new deep and fast package C-states, likes \tech, for future server CPUs targeting datacenters running microservices.


\section*{Acknowledgments}
\begin{sloppypar}
This project has received funding from the European Union’s Horizon 2020 research and innovation programme under the Marie Skłodowska-Curie grant agreement No 101029391.
\end{sloppypar}


\bibliographystyle{IEEEtranSN}
\bibliography{refs}

\end{document}